\documentclass[sigconf,balance=false]{acmart}

\usepackage{popets}

\copyrightyear{2024}
\acmYear{2024}
\acmVolume{2024}
\acmNumber{X}
\acmDOI{XXXXXXX.XXXXXXX}
\acmISBN{}
\acmConference{}
\usepackage{tikz}
\usepackage{amsmath}
\usepackage{tabularx}
\usepackage{multirow}
\usepackage{comment}
\usepackage[colorinlistoftodos]{todonotes}

\usepackage{filecontents}

\begin{document}


\title{Empirical Analysis of Sri Lankan Mobile Health Ecosystem: \\ A Precursor to an Effective Stakeholder Engagement}

 \author{Kenneth Thilakarathna,$^{1, 2}$ Sachintha Pitigala,$^{2}$ Jayantha Fernando,$^{3}$ Primal Wijesekera,$^{4,5}$}
 \affiliation{%
   \institution{$^1$University of Colombo School of Computing, $^2$University of Kelaniya, $^3$Heritage Partners, $^4$ICSI, $^5$UC Berkeley}
   \country{}
 }

\begin{abstract}

Sri Lanka recently passed its first privacy legislation covering a wide range of sectors, including health. As a precursor for effective stakeholder engagement in the health domain to understand the most effective way to implement legislation in healthcare, we have analyzed 41 popular mobile apps and web portals. We found that 78\% of the tested systems have third-party domains receiving sensitive health data with minimal visibility to the consumers. We discuss how this will create potential issues in preparing for the new privacy legislation.

\end{abstract}
 
\keywords{privacy, health surveillance, compliance}





\maketitle


\newenvironment{packed_enum}{
\begin{enumerate}
  \setlength{\itemsep}{1pt}
  \setlength{\parskip}{0pt}
  \setlength{\parsep}{0pt}
}{\end{enumerate}}

\newenvironment{packed_item}{
\begin{itemize}
  \setlength{\itemsep}{1pt}
  \setlength{\parskip}{0pt}
  \setlength{\parsep}{0pt}
}{\end{itemize}}

\vspace*{-0.7\baselineskip}

\section{Introduction}
\label{intro}

Sri Lanka's mature and effective public health ecosystem has proven equally efficient compared to major economies. Given the open and unrestricted nature of the health ecosystem, patients can consult any medical practitioner at will without worrying about insurance or the cost. Inadvertently, this has also created a vibrant digital ecosystem where consumers in Sri Lanka have long enjoyed the comfort of scheduling appointments over the phone digitally.

As the first in the region, Sri Lanka recently passed its first nationwide privacy legislation~\cite{pdpa}. Europe and the US, despite having regulations in place for some time, are facing continuous threats of breaching consumer privacy expectations and regulatory violations. Research at the intersection of Tech, Policy, and Legal has shown various reasons, such as lack of awareness, miscomprehension of the regulatory requirements, miscommunication, the lack of transparency in the third-party code and systems used by developers, and finally, lack of a strong institutional framework due to budgetary and bureaucratic constraints. This long list of reasons has produced the notion of law in the books vs. law in the code.

Sri Lanka is uniquely placed in this journey since the country is at an early phase of setting up the institutional framework for the implementation of privacy legislation enacted recently. The core objective of the work is to utilize the golden period and ensure that relevant authorities and stakeholders in the health domain are properly informed by empirical evidence and aware of their responsibilities towards a privacy-conscious ecosystem.

\section{Related Work}
\label{related}

Prior studies have investigated potential regulatory violations in mobile apps in the US~\cite{nguyen2021share, reyes2018won, reyes2018won, han2020price, feal2020angel, samarin2023lessons} while another line of work examines the compliance of mobile health apps in the context of GDPR~\cite{Fan2020gdprcompliance, Mulder2019HealthAppsPrivacyGDPR, Benjumea2019gdprcompliance, Papageorgiou2018gdprcompliance}. The focus of privacy studies has delve into the healthcare ecosystem through multiple subsystems such as analyzing privacy of femtech health~\cite{shipp2020private, epstein2017examining, rosas2019future}, mobile-based COVID-19 tracing applications for their privacy implications~\cite{chidambaram2020observational, simmhan2020gocoronago},  operation of mobile therapy apps, a subset of the telehealth app ecosystem~\cite{iwaya2023privacy, hassan2023unveiling} to name a few. Another research trajectory assesses the availability, scope, and transparency of mobile health app privacy policies~\cite{Benjumea2020AssessmentFairness, Sunyaev2014MobileHealthPrivacyPolicy, Sampat2017PrivacyRisks, OLOUGHLIN2019110} where there are studies focused on understanding the developers' role in health privacy and challenges faced in producing complied software solutions~\cite{tahaei2022privacy, bednar2019engineering, hadar2018privacy, Mustafa2019devRoleInCompliance, Muchagata2018DevRoleInCompliance, aljedaani2020empirical, aljedaani2021challenges}. In the US, legal authorities such as the FTC and HHS scrutinize mobile health apps for compliance~\cite{ftc_letters, betterhelp, premom, goodrx}. 
FTC has used the HBNR~\cite{hbnr} and FTC Act~\cite{FTCAct} to pursue these actions as these apps traditionally fall outside the scope of HIPAA~\cite{hipaa_ref}.

\section{Background}
\label{background}

\subsection{What is a health app?}

Definitions of "health app," "medical app," or "wellness app" can overlap, and there is no universally accepted definition for these terms. For the context of this work, we define a health app as an Android-based mobile app or a mobile-ready website that lets consumers (patients) communicate with a healthcare provider and schedule an appointment or give specific information or guidance on a specific health condition such as diabetes, or online pharmacy. 


\subsection{Applicable Privacy Legislation}

In 2022, the government of Sri Lanka passed its Personal Data Protection Act, No. 9 of 2022 (PDPA), which is a commendable starting point for a privacy ecosystem~\cite{pdpa}. PDPA is a comprehensive privacy legislation such as CCPA or GDPR; however, Health data is defined under a special category with extra protection.

\section{Methodology}
\label{methodology}

\subsection{Mobile Analysis Tool}

We focused our analysis on the Android ecosystem, given it is the most prevalent mobile operating system~\cite{androidshare} and our in-house tools on Android.
In our custom Android platform (based on v9.0-\_r39), we modified the platform to enable the real-time monitoring of apps' access to protected resources (e.g., location data, address book contacts, etc.) and sharing over the network. We have also implemented stringent guards to hide our instrumentation from the latest array of anti analysis techniques~\cite{lim2018survey, Zhang_15, Talsec, Appsealing, Promon}.

Our network interception occurs at two different points: one at the default Android network stack at \textit{conscrypt} library level just before the SSL\_read and SSL\_write, and one at the webview library level intercepting XHR requests. Our instrumentation does not have any false positives, but there is a probability of false negatives due to the coverage in app execution -- the data should be treated as the absolute lower bound of what is happening in the real world.

\vspace*{-0.7\baselineskip}

\subsection{App Corpus}

We looked for Android apps and websites that let a patient make an appointment with a physician, provide information or guidance on specific conditions, online pharmacy systems, rehabilitation apps, or sites for leading hospitals in the Sri Lanka Google Play Store from 01 June 2024 until 07 June 2024 and in the Google website. We ended up testing 41 Android-based apps and websites. The dataset includes 5 Android apps and 37 websites accessed through Android: 11 hospitals, 14 medical clinics, four information sites targeting specific conditions, seven online pharmacies, and five physician scheduling portals. The data set also has different types of health apps focusing on fertility, diabetes, cancer, mental health, etc. From here onward, we refer to both apps and websites as health systems.

\subsection{Testing Procedure} We tested the apps and websites using Pixel 3a phones running our instrumented Android. We explored the apps/websites as much as possible go over as many options and links as possible. Whenever possible, we also searched for specific conditions or a specific physician to see whether such information is shared with third-parties.

Android offers more sensitive resources and functionalities than the desktop environment. Hence, understanding how these websites behave in the mobile ecosystem is important. During the app testing, the researcher recorded all the sensitive data used, such as synthetic names, all the synthetic health data used to fill the questions, sensitive pages visited, such as "help on anxiety." Once the network traces are decoded, we use a script to look for any transaction with specific strings used in the testing.  

\subsection{Health Data}

The definition of health data under PDPA~\cite{pdpa} is ''personal data
related to the physical or psychological health of a natural person, which includes any information that indicates his health situation or status;;'' and HHS recently issued a guidance~\cite{hhs-2022}, identifying that even sharing the app name along with a unique ID of the consumer would fall within the definition of Personal Health Information (PHI), as it identifies the individual's past, present or future health or health care or payment. Keeping this new legal principles in mind, we labeled the following items as health data as any of the following can be used to infer their health conditions or health interests: conventional health data, app usage (apps targeting specific medical conditions), search queries within health apps (searching for a medical condition, a medication, etc.), navigation with the app (viewing pages focusing on a specific medical condition).

\subsection{Ethical Considerations}

Based on our number of prior conversations with our Institutional Review Board on large-scale mobile app analysis for compliance, we determined we do not require an IRB review for this study. We are not examining any human subject but only the app execution.

\subsection*{Ecological Validity}

Android apps are likely to detect the geolocation based on the IP and might change their behavior accordingly. All Sri Lankan based apps were executed in Sri Lanka to preserve the ecological validity of the test environment. 
We only have one app that stopped its execution after detecting the underlying custom Android OS.


\section{Data Sharing Practices}

\begin{table}[]
\resizebox{\columnwidth}{!}{%
\begin{tabular}{|l|l|c|l|}
\hline
\multicolumn{1}{|c|}{\textbf{\begin{tabular}[c]{@{}c@{}}Third-Party \\ Recepient\end{tabular}}} & \multicolumn{1}{c|}{\textbf{Type}} & \textbf{\begin{tabular}[c]{@{}c@{}}Number of\\ Apps\end{tabular}} & \multicolumn{1}{c|}{\textbf{\begin{tabular}[c]{@{}c@{}}Health Data\\ Status\end{tabular}}} \\ \hline
\begin{tabular}[c]{@{}l@{}}Google \\ Analytics\end{tabular} & \begin{tabular}[c]{@{}l@{}}Tracking/\\ Event Reporting\end{tabular} & 10 & Non-compliant \\ \hline
Google Search & Search & 8 & \begin{tabular}[c]{@{}l@{}}Likely \\ non-compliant\end{tabular} \\ \hline
Facebook & \begin{tabular}[c]{@{}l@{}}Tracking/\\ Event Reporting\end{tabular} & 6 & Non-compliant \\ \hline
Doubleclick & Advertising & 5 & Non-compliant \\ \hline
\end{tabular}%
}
\caption{Third-party recipients of physician information. }
\label{physician}
\vspace*{-1.5\baselineskip}
\end{table}

\begin{table}[]
\resizebox{\columnwidth}{!}{%
\begin{tabular}{|l|l|c|l|}
\hline
\multicolumn{1}{|c|}{\textbf{\begin{tabular}[c]{@{}c@{}}Third-Party \\ Recepient\end{tabular}}} & \multicolumn{1}{c|}{\textbf{Type}} & \textbf{\begin{tabular}[c]{@{}c@{}}Number of\\ Apps\end{tabular}} & \multicolumn{1}{c|}{\textbf{\begin{tabular}[c]{@{}c@{}}Health Data\\ Status\end{tabular}}} \\ \hline
\begin{tabular}[c]{@{}l@{}}Google \\ Analytics\end{tabular} & \begin{tabular}[c]{@{}l@{}}Tracking/\\ Event Reporting\end{tabular} & 24 & Non-compliant \\ \hline
Google Search & Search & 11 & \begin{tabular}[c]{@{}l@{}}Likely \\ non-compliant\end{tabular} \\ \hline
Facebook & \begin{tabular}[c]{@{}l@{}}Tracking/\\ Event Reporting\end{tabular} & 5 & Non-compliant \\ \hline
Zoho & CDN & 5 & Unknown \\ \hline
Doubleclick & Advertising & 4 & Non-compliant \\ \hline
GStatic & CDN & 3 & Unknown \\ \hline
\begin{tabular}[c]{@{}l@{}}Google\\ Tag Manager\end{tabular} & Tracking & 2 & Non-compliant \\ \hline
Cloudfare & CDN & 2 & Unknown \\ \hline
Private Site & Personal & 1 & \begin{tabular}[c]{@{}l@{}}Likely \\ non-compliant\end{tabular} \\ \hline
\end{tabular}%
}
\caption{Recipients of sensitive usage information. }
\label{attribution}
\vspace*{-1.7\baselineskip}
\end{table}

\paragraph{Physician Information}: Since the pandemic, systems allowing scheduling appointments online have soared, and most of them went unchecked until recently. Online physician scheduling is quite popular in Sri Lanka. In the health compliance ecosystem, they play a crucial role. Our dataset has seventeen systems that allow patients to search for physicians, get scheduling information, or make an appointment directly. We have observed that ten health systems (58.82\% out of 17 health systems with the scheduling feature) have shared sensitive physician information with third parties. Physician information can divulge a patient's condition as a highly sensitive data point. Table~\ref{physician} lists all the third-party recipients of such sensitive information. The last column denotes whether the respective recipient will treat the health data without violating any regulatory or consumer privacy expectations.

\paragraph{Usage Information}: Event reporting in online systems is usually harmless for the consumers. But, in a health system, it can diffuse sensitive information such as potential interest or a condition a patient has. For example, a repeated visit to infertility pages can likely expose consumer's highly sensitive condition. Hence, sharing usage information should be done carefully with masking. This concern is confirmed by HSS (in the US) as per their latest guidance on regulatory expectations~\cite{hhs-2022}. 


In our dataset, we have 27 (65.85\% of our test-pool) apps sharing highly sensitive usage information with third parties who are likely to use such information for user tracking and profiling. This usage information includes visits to infertility treatment pages, mental illness tests, physician pages specializing in specific conditions. In an ideal setup, patients are highly unlikely to share their interests and why they visit such pages. Table~\ref{attribution} lists all the third-party recipients of such sensitive information. The last column denotes whether the recipient will treat the health data without violating regulatory or consumer privacy expectations -- this is determined based on their public documents.

We observed that nine health systems shared the search queries we used during the testing with third-party recipients—Google Analytics (8 apps) and Facebook (2 apps) are the most common recipients. Similar to app usage, search queries are sensitive, such as physicians, symptoms, and a particular medicine, all of which could expose sensitive conditions associated with the patient.

We also observed one health system sending sensitive health information over the Internet unencrypted, jeopardizing the confidentiality and integrity of the patients' health data.

Out of the 33 health systems sharing data with third parties, only nine health systems (27.27\%) have acknowledged third-party data sharing in their privacy policies, and, overall, 27 (~65\% of our test-pool) health systems did not have a privacy policy.

Except for two Android apps (which shared AAID with Facebook and Google Analytics), none of the other systems shared soft or persistent IDs with third parties. From a developer perspective, no sensitive health data shared had an ID linking to the patient. The biggest caveat, however, is the patient's IP address. Even HIPAA labels the IP address as one of the eighteen HIPAA identifiers that can be used to link to a specific person. The extent of the linkage depends on the nature of the mobile phone's connection. If it is a home WIFI, it is easy to link each data transfer to a specific person along with the IP address and many other meta information.

\vspace*{-0.7\baselineskip}

\section{Regulatory Preparation}

The main objective of this work is to understand the current status of the mobile health ecosystem as a precursor to understanding the cost, responsibilities, and challenges faced by developers and health organizations in complying with the new Privacy legislation.

The preparation has to be done in two ways: by managing consent and by providing data subject rights. We recently conducted a  focus group to understand stakeholder perspectives on the new privacy legislation~\cite{roundtable}. Participants raised both of these tasks as sources of cost in the process of complying with new legislation.

None of the health systems we analyzed have any sort of consent management (except for one website that had cookie consent). This will be one of the first major changes for health systems to properly manage consumers' consent. There are several challenges to effectively obtaining an \textit{informed} consent from patients. 

PDPA requires the controller to properly convey one or more predefined purposes to the data subject before obtaining consent unless exempted under PDPA guidelines such as legal obligations. Given the widespread use of third parties receiving sensitive health data, it will be a challenge for health systems to set a predefined purpose properly. Especially once a data controller shares data with the likes of Facebook and Google Analytics, it is hard to dictate how they are going to use the health data.

Another major change would be to obtain consent for cross-border data transfer. All third-party data recipients are not based in Sri Lanka; hence, as per PDPA clauses, health systems need to obtain consent from the users of the health systems properly.

The heavy use of third-party trackers such as Google Analytics, Facebook, and DoubleClick further complicates since controllers need to properly expose how the data recipients are going to profile customers based on health data. Apart from the knowledge that most of such data is used in Advertising, it is a black box for outsiders to understand how the whole mobile ad ecosystem behaves, leaving the data controllers in Sri Lanka in the dark. PDPA has a separate clause for profiling, especially when using special category data such as health.

PDPA (similar to the clauses in GDPR and CCPA) emphasizes providing data subjects (patients in this context) with an array of rights: access, erase, and withdraw consent. Our prior work on implementation of data subject requests (DSR) in CCPA~\cite{samarin2023lessons, samarin2023understanding} showed controllers have a hard time accurately responding to DSR because controllers are not fully aware of how third parties collected data from data subjects while executing within the controller's app. Given the widespread use of third-party trackers, Sri Lankan health systems will face the same issue.

PDPA sets clear guidelines on what information should be available for the consumer for transparency. Privacy policy is one of the key techniques to properly convey data practices, purposes, and other relevant information. Most health systems in Sri Lanka do not have any privacy policy. We believe this will be an easy first step for many organizations to publish an accurate privacy policy. Still, this step will also be affected by third-party data collection's opaque nature and their purposes.

Literature has looked into how developers understand their regulatory responsibilities and the gaps in their comprehension of the regulations~\cite{alomar2022developers}. Further work is needed to understand why developers share health data with third parties that have publicly asked not to share heath data with them~\footnote{https://support.google.com/analytics/answer/13297105?hl=en}~\footnote{https://web.facebook.com/business/help/361948878201809?id=188852726110565\&\_rdc=1\&\_rdr}. Literature has proposed solutions such as the use of SBOM to communicate compliance restrictions~\cite{wijesekers24_conpro}. This is one of the overarching objectives of this work, i.e., to work with stakeholders to understand how to effectively implement privacy legislation while helping the likes of developers of health systems providing proper guidance and tools. 

The Data Protection Authority of Sri Lanka is keen to work with stakeholders to understand their perspectives and figure out the best way to roll out the implementation with the help of relevant parties such as health organizations, patients, developers of health systems, and legal practitioners. We have already conducted one stakeholder engagement to understand the cost, challenges, and opportunities that lie ahead in the compliance process. We hope this work will set an effective precursor for engaging with professionals in the healthcare domain to figure out their costs and challenges in preparing for the new legislation. The objective of this is not to blame anyone but to figure out where the help is most needed.

\begin{acks}
This work was supported by the U.S. National Science Foundation (under grant CNS-2055772 \& CNS-2217771 ). 
\end{acks}

\bibliographystyle{plain}
\bibliography{main}

\begin{thebibliography}{10}

\bibitem{hipaa_ref}
Coppa support.
\newblock \url{https://github.com/prebid/prebid.github.io/pull/3476}.
\newblock Accessed: 2022-01-28.

\bibitem{FTCAct}
{Federal Trade Commission (FTC) Act}.
\newblock 15 U.S.C. \S45(a)(1).

\bibitem{pdpa}
Personal data protection act.
\newblock \url{https://www.parliament.lk/uploads/acts/gbills/english/6242.pdf}.

\bibitem{roundtable}
Round table discussion on pdpa.
\newblock \url{https://www.linkedin.com/pulse/roundtable-discussion-privacy-regulation-beyond-primal-wijesekera-lor3c/}.

\bibitem{aljedaani2021challenges}
B~Aljedaani and MA~Babar.
\newblock Challenges with developing secure mobile health applications: Systematic review. jmir mhealth uhealth 9, e15654, 2021.

\bibitem{aljedaani2020empirical}
Bakheet Aljedaani, Aakash Ahmad, Mansooreh Zahedi, and M~Ali Babar.
\newblock An empirical study on developing secure mobile health apps: The developers' perspective.
\newblock In {\em 2020 27th Asia-Pacific Software Engineering Conference (APSEC)}, pages 208--217. IEEE, 2020.

\bibitem{alomar2022developers}
Noura Alomar and Serge Egelman.
\newblock Developers say the darnedest things: Privacy compliance processes followed by developers of child-directed apps.
\newblock {\em Proceedings on Privacy Enhancing Technologies}, 4(2022):24, 2022.

\bibitem{bednar2019engineering}
Kathrin Bednar, Sarah Spiekermann, and Marc Langheinrich.
\newblock Engineering privacy by design: Are engineers ready to live up to the challenge?
\newblock {\em The Information Society}, 35(3):122--142, 2019.

\bibitem{Benjumea2020AssessmentFairness}
J~Benjumea, J~Ropero, O~Rivera-Romero, E~Dorronzoro-Zubiete, and A~Carrasco.
\newblock Assessment of the fairness of privacy policies of mobile health apps: Scale development and evaluation in cancer apps.
\newblock {\em JMIR Mhealth Uhealth}, 8(7):e17134, 2020.

\bibitem{Benjumea2019gdprcompliance}
Jaime Benjumea, Enrique Dorronzoro, Jorge Ropero, Octavio Rivera-Romero, and Alejandro Carrasco.
\newblock Privacy in mobile health applications for breast cancer patients.
\newblock In {\em 2019 IEEE 32nd International Symposium on Computer-Based Medical Systems (CBMS)}, pages 634--639, 2019.

\bibitem{chidambaram2020observational}
Swathikan Chidambaram, Simon Erridge, James Kinross, and Sanjay Purkayastha.
\newblock Observational study of uk mobile health apps for covid-19.
\newblock {\em The Lancet Digital Health}, 2(8):e388--e390, 2020.

\bibitem{hbnr}
Federal~Trade Commission.
\newblock Health breach notification rule.
\newblock \url{https://www.ftc.gov/legal-library/browse/rules/health-breach-notification-rule/}.
\newblock Accessed: 2023-10-15.

\bibitem{androidshare}
Stat Counter.
\newblock Mobile operating system market share worldwide.
\newblock https://gs.statcounter.com/os-market-share/mobile/worldwide, 2023.
\newblock Accessed: 2023-10-15.

\bibitem{epstein2017examining}
Daniel~A Epstein, Nicole~B Lee, Jennifer~H Kang, Elena Agapie, Jessica Schroeder, Laura~R Pina, James Fogarty, Julie~A Kientz, and Sean Munson.
\newblock Examining menstrual tracking to inform the design of personal informatics tools.
\newblock In {\em Proceedings of the 2017 CHI Conference on Human Factors in Computing Systems}, pages 6876--6888, 2017.

\bibitem{goodrx}
Leselye Fair.
\newblock First ftc health breach notification rule case addresses goodrx’s not-so-good privacy practices.
\newblock \url{https://www.ftc.gov/business-guidance/blog/2023/02/first-ftc-health-breach-notification-rule-case-addresses-goodrxs-not-so-good-privacy-practices}, 2023.
\newblock Accessed: 2023-10-15.

\bibitem{Fan2020gdprcompliance}
Ming Fan, Le~Yu, Sen Chen, Hao Zhou, Xiapu Luo, Shuyue Li, Yang Liu, Jun Liu, and Ting Liu.
\newblock An empirical evaluation of gdpr compliance violations in android mhealth apps.
\newblock In {\em 2020 IEEE 31st International Symposium on Software Reliability Engineering (ISSRE)}, pages 253--264, 2020.

\bibitem{feal2020angel}
{\'A}lvaro Feal, Paolo Calciati, Narseo Vallina-Rodriguez, Carmela Troncoso, Alessandra Gorla, et~al.
\newblock Angel or devil? a privacy study of mobile parental control apps.
\newblock {\em Proceedings on Privacy Enhancing Technologies (PoPETs)}, 2020(2):314--335, 2020.

\bibitem{betterhelp}
FTC.
\newblock Betterhelp, inc., in the matter of.
\newblock \url{https://www.ftc.gov/legal-library/browse/cases-proceedings/2023169-betterhelp-inc-matter}, 2023.
\newblock Accessed: 2023-10-15.

\bibitem{premom}
FTC.
\newblock Ovulation tracking app premom will be barred from sharing health data for advertising under proposed ftc order.
\newblock \url{https://www.ftc.gov/news-events/news/press-releases/2023/05/ovulation-tracking-app-premom-will-be-barred-sharing-health-data-advertising-under-proposed-ftc}, 2023.
\newblock Accessed: 2023-10-15.

\bibitem{ftc_letters}
HHS~\& FTC.
\newblock {Use of Online Tracking Technologies}.
\newblock \url{https://www.hhs.gov/sites/default/files/ocr-ftc-letters-re-use-online-tracking-technologies.pdf}, July 20 2023.

\bibitem{hadar2018privacy}
Irit Hadar, Tomer Hasson, Oshrat Ayalon, Eran Toch, Michael Birnhack, Sofia Sherman, and Arod Balissa.
\newblock Privacy by designers: software developers’ privacy mindset.
\newblock {\em Empirical Software Engineering}, 23:259--289, 2018.

\bibitem{han2020price}
Catherine Han, Irwin Reyes, {\'A}lvaro Feal, Joel Reardon, Primal Wijesekera, Narseo Vallina-Rodriguez, Amit Elazar, Kenneth~A Bamberger, and Serge Egelman.
\newblock The price is (not) right: Comparing privacy in free and paid apps.
\newblock {\em Proceedings on Privacy Enhancing Technologies (PoPETs)}, 2020(3):222--242, 2020.

\bibitem{hassan2023unveiling}
Muhammad Hassan and Masooda Bashir.
\newblock Unveiling privacy measures in mental health applications.
\newblock In {\em Adjunct Proceedings of the 2023 ACM International Joint Conference on Pervasive and Ubiquitous Computing \& the 2023 ACM International Symposium on Wearable Computing}, pages 648--654, 2023.

\bibitem{hhs-2022}
Health and Human Services.
\newblock {Use of Online Tracking Technologies by HIPAA Covered Entities and Business Associates}.
\newblock \url{https://www.hhs.gov/hipaa/for-professionals/privacy/guidance/hipaa-online-tracking/index.html}, December 01 2022.

\bibitem{Appsealing}
{INKA Entworks Inc.}
\newblock Appsealing.
\newblock \url{https://www.appsealing.com/}, September 30 2021.

\bibitem{iwaya2023privacy}
Leonardo~Horn Iwaya, M~Ali Babar, Awais Rashid, and Chamila Wijayarathna.
\newblock On the privacy of mental health apps: An empirical investigation and its implications for app development.
\newblock {\em Empirical Software Engineering}, 28(1):2, 2023.

\bibitem{lim2018survey}
Jongsu Lim, Yonggu Shin, Sunjun Lee, Kyuho Kim, and Jeong~Hyun Yi.
\newblock Survey of dynamic anti-analysis schemes for mobile malware.
\newblock {\em J. Wirel. Mob. Networks Ubiquitous Comput. Dependable Appl.}, 9(3):39--49, 2018.

\bibitem{Muchagata2018DevRoleInCompliance}
Joana Muchagata and Ana Ferreira.
\newblock Translating gdpr into the mhealth practice.
\newblock In {\em 2018 International Carnahan Conference on Security Technology (ICCST)}, pages 1--5, 2018.

\bibitem{Mulder2019HealthAppsPrivacyGDPR}
Trix Mulder.
\newblock Health apps, their privacy policies and the gdpr.
\newblock {\em European Journal of Law and Technology}, Jun 2019.
\newblock University of Groningen Faculty of Law Research Paper No. 15/2020.

\bibitem{Mustafa2019devRoleInCompliance}
Uzma Mustafa, Eckhard Pflugel, and Nada Philip.
\newblock A novel privacy framework for secure m-health applications: The case of the gdpr.
\newblock In {\em 2019 IEEE 12th International Conference on Global Security, Safety and Sustainability (ICGS3)}, pages 1--9, 2019.

\bibitem{nguyen2021share}
Trung~Tin Nguyen, Michael Backes, Ninja Marnau, and Ben Stock.
\newblock Share first, ask later (or never?) studying violations of {GDPR's} explicit consent in android apps.
\newblock In {\em 30th USENIX Security Symposium (USENIX Security 21)}, pages 3667--3684, Berkeley, CA, USA, 2021. USENIX.

\bibitem{OLOUGHLIN2019110}
Kristen O'Loughlin, Martha Neary, Elizabeth~C. Adkins, and Stephen~M. Schueller.
\newblock Reviewing the data security and privacy policies of mobile apps for depression.
\newblock {\em Internet Interventions}, 15:110--115, 2019.

\bibitem{Papageorgiou2018gdprcompliance}
Achilleas Papageorgiou, Michael Strigkos, Eugenia Politou, Efthimios Alepis, Agusti Solanas, and Constantinos Patsakis.
\newblock Security and privacy analysis of mobile health applications: The alarming state of practice.
\newblock {\em IEEE Access}, 6:9390--9403, 2018.

\bibitem{Promon}
Promon.
\newblock {Shield: In-App Protection and Security for Mobile Apps}.
\newblock \url{https://promon.co/products/mobile-app-protection/}, September 30 2021.

\bibitem{reyes2018won}
Irwin Reyes, Primal Wijesekera, Joel Reardon, Amit Elazari Bar~On, Abbas Razaghpanah, Narseo Vallina-Rodriguez, Serge Egelman, et~al.
\newblock ``won't somebody think of the children?'' examining coppa compliance at scale.
\newblock {\em Proceedings on Privacy Enhancing Technologies (PoPETs)}, 2018(3):63--83, 2018.

\bibitem{rosas2019future}
Celia Rosas.
\newblock The future is femtech: Privacy and data security issues surrounding femtech applications.
\newblock {\em Hastings Bus. LJ}, 15:319, 2019.

\bibitem{samarin2023lessons}
Nikita Samarin, Shayna Kothari, Zaina Siyed, Oscar Bjorkman, Reena Yuan, Primal Wijesekera, Noura Alomar, Jordan Fischer, Chris Hoofnagle, and Serge Egelman.
\newblock Lessons in vcr repair: Compliance of android app developers with the california consumer privacy act (ccpa).
\newblock {\em arXiv preprint arXiv:2304.00944}, 2023.

\bibitem{samarin2023understanding}
Nikita Samarin and Primal Wijesekera.
\newblock Understanding how third-party libraries in mobile apps affect responses to subject access requests.
\newblock 2023.

\bibitem{Sampat2017PrivacyRisks}
Brinda~Hansraj Sampat and Bala Prabhakar.
\newblock Privacy risks and security threats in mhealth apps.
\newblock {\em Journal of International Technology and Information Management}, 26(4), 2017.

\bibitem{shipp2020private}
Laura Shipp and Jorge Blasco.
\newblock How private is your period?: A systematic analysis of menstrual app privacy policies.
\newblock {\em Proc. Priv. Enhancing Technol.}, 2020(4):491--510, 2020.

\bibitem{simmhan2020gocoronago}
Yogesh Simmhan, Tarun Rambha, Aakash Khochare, Shriram Ramesh, Animesh Baranawal, John~Varghese George, Rahul~Atul Bhope, Amrita Namtirtha, Amritha Sundararajan, Sharath~Suresh Bhargav, et~al.
\newblock Gocoronago: privacy respecting contact tracing for covid-19 management.
\newblock {\em Journal of the Indian Institute of Science}, 100:623--646, 2020.

\bibitem{Sunyaev2014MobileHealthPrivacyPolicy}
Ali Sunyaev, Tobias Dehling, Patrick~L Taylor, and Kenneth~D Mandl.
\newblock {Availability and quality of mobile health app privacy policies}.
\newblock {\em Journal of the American Medical Informatics Association}, 22(e1):e28--e33, 08 2014.

\bibitem{tahaei2022privacy}
Mohammad Tahaei, Julia Bernd, and Awais Rashid.
\newblock Privacy, permissions, and the health app ecosystem: A stack overflow exploration.
\newblock In {\em Proceedings of the 2022 European Symposium on Usable Security}, pages 117--130, 2022.

\bibitem{Talsec}
Talsec.
\newblock {Talsec freeRASP}.
\newblock \url{https://www.talsec.app/freerasp-in-app-protection-security-talsec}, September 30 2021.

\bibitem{wijesekers24_conpro}
Primal Wijesekera.
\newblock Health compliance through a transparent supply chain.
\newblock 2024.

\bibitem{Zhang_15}
Hang Zhang, Dongdong She, and Zhiyun Qian.
\newblock Android root and its providers: A double-edged sword.
\newblock In {\em Proceedings of the 22nd ACM SIGSAC Conference on Computer and Communications Security}, CCS '15, pages 1093--1104, New York, NY, USA, 2015. Association for Computing Machinery.

\end{thebibliography}

\begin{table*}[]
\begin{tabular}{|l|l|l|l|l|}
\hline
\multicolumn{1}{|c|}{\textbf{\begin{tabular}[c]{@{}c@{}}ID / URL\end{tabular}}} & \multicolumn{1}{c|}{\textbf{App Name}} & \textbf{\begin{tabular}[c]{@{}c@{}}App Type \end{tabular}} & \multicolumn{1}{c|}{\textbf{\begin{tabular}[c]{@{}c@{}}Service Category \end{tabular}}} & \multicolumn{1}{c|}{\textbf{Owner Category}}\\ \hline

\begin{tabular}[c]{@{}l@{}}com.inova.velocity\end{tabular} & \begin{tabular}[c]{@{}l@{}}Doc990\end{tabular} & Android App & Physician Appointment  & Third-Party/Telecom \\ \hline
\begin{tabular}[c]{@{}l@{}}doc.lk\end{tabular} & \begin{tabular}[c]{@{}l@{}}Doc990\end{tabular} & Website & Physician Appointment  & Third-Party/Telecom \\ \hline
\begin{tabular}[c]{@{}l@{}}com.developer.odoc\end{tabular} & \begin{tabular}[c]{@{}l@{}}oDoc - \\ Video Consultations\end{tabular} & Android App & Video Consultation  & First-party/technical \\ \hline

\begin{tabular}[c]{@{}l@{}}asirihealth.com\end{tabular} & \begin{tabular}[c]{@{}l@{}}Asiri Health\end{tabular} & Website & Appointment/generic  & Hospital \\ \hline
\begin{tabular}[c]{@{}l@{}}www.durdans.com/appointments/\end{tabular} & \begin{tabular}[c]{@{}l@{}}Durdans\end{tabular} & Website & Appointment/generic  & Hospital \\ \hline
\begin{tabular}[c]{@{}l@{}}www.nawaloka.com/channeling/\end{tabular} & \begin{tabular}[c]{@{}l@{}}Nawaloka\end{tabular} & Website & Generic  & Hospital \\ \hline
\begin{tabular}[c]{@{}l@{}}www.lankahospitals.com/en/doctor-channeling/\end{tabular} & \begin{tabular}[c]{@{}l@{}}Lanka Hospital\end{tabular} & Website & Generic  & Hospital \\ \hline
\begin{tabular}[c]{@{}l@{}}www.delmonhospital.com/en/find-a-doctor\end{tabular} & \begin{tabular}[c]{@{}l@{}}Delmon Hospital\end{tabular} & Website & Generic  & Hospital \\ \hline
\begin{tabular}[c]{@{}l@{}}www.ninewellshospital.lk/appointment-booking/\end{tabular} & \begin{tabular}[c]{@{}l@{}}Nine Wells\end{tabular} & Website & Appointment/generic  & Hospital \\ \hline
\begin{tabular}[c]{@{}l@{}}https://medihelphealth.com/ \\ https://www.medihelp.lk/\end{tabular} & \begin{tabular}[c]{@{}l@{}}Medihelp\end{tabular} & Website & Appointment/generic  & Hospital \\ \hline
\begin{tabular}[c]{@{}l@{}}www.wishfertility.lk/\end{tabular} & \begin{tabular}[c]{@{}l@{}}Wish Fertility\end{tabular} & Website & Generic  & Medical Services \\ \hline
\begin{tabular}[c]{@{}l@{}}hemashospitals.com/\end{tabular} & \begin{tabular}[c]{@{}l@{}}Hemas Hospitals\end{tabular} & Website & Generic  & Hospital \\ \hline
\begin{tabular}[c]{@{}l@{}}ivflanka.com/\end{tabular} & \begin{tabular}[c]{@{}l@{}}IVF Lanka\end{tabular} & Website & Generic  & Medical Services \\ \hline
\begin{tabular}[c]{@{}l@{}}www.suvika.lk/\end{tabular} & \begin{tabular}[c]{@{}l@{}}Suvika\end{tabular} & Website & Generic  & Medical Services \\ \hline
\begin{tabular}[c]{@{}l@{}}www.pfrcivf.lk/\end{tabular} & \begin{tabular}[c]{@{}l@{}}PCR IVF\end{tabular} & Website & Generic  & Medical Services \\ \hline
\begin{tabular}[c]{@{}l@{}}jeewakaprivatehospital.lifegroup.lk/\\channel-your-doctor/\end{tabular} & \begin{tabular}[c]{@{}l@{}}Jeewaka Hospital\end{tabular} & Website & Generic  & Hospital \\ \hline
\begin{tabular}[c]{@{}l@{}}www.goldenkeyhospitals.com/\\ask-online-from-a-consultant\end{tabular} & \begin{tabular}[c]{@{}l@{}}Golden EENT\end{tabular} & Website & Generic  & Medical Services \\ \hline
\begin{tabular}[c]{@{}l@{}}www.inspirationscare.lk/\#\end{tabular} & \begin{tabular}[c]{@{}l@{}}Inspiration\end{tabular} & Website & Generic  & Medical Services \\ \hline
\begin{tabular}[c]{@{}l@{}}santadorahospital.com/\end{tabular} & \begin{tabular}[c]{@{}l@{}}Santa Dora\end{tabular} & Website & Generic  & Medical Services \\ \hline
\begin{tabular}[c]{@{}l@{}}arogyahospitals.lk/\end{tabular} & \begin{tabular}[c]{@{}l@{}}Arogya\end{tabular} & Website & Generic  & Medical Services \\ \hline
\begin{tabular}[c]{@{}l@{}}www.diabetessrilanka.org/\end{tabular} & \begin{tabular}[c]{@{}l@{}}National Diabetes \\Centre\end{tabular} & Website & Generic  & Medical Info Services \\ \hline
\begin{tabular}[c]{@{}l@{}}www.csth.health.gov.lk/\end{tabular} & \begin{tabular}[c]{@{}l@{}}CSTH\end{tabular} & Website & Generic  & Medical Info Services \\ \hline
\begin{tabular}[c]{@{}l@{}}www.jaffnadiabeticcentre.org/\end{tabular} & \begin{tabular}[c]{@{}l@{}}Jaffna Diabetic \\Centre\end{tabular} & Website & Generic  & Medical Info Services \\ \hline
\begin{tabular}[c]{@{}l@{}}www.nawinna.com/\end{tabular} & \begin{tabular}[c]{@{}l@{}}Nawinna\end{tabular} & Website & Generic  & Medical Services \\ \hline
\begin{tabular}[c]{@{}l@{}}www.queensburyhospitals.lk\end{tabular} & \begin{tabular}[c]{@{}l@{}}Queensbury\end{tabular} & Website & Generic  & Hospital \\ \hline
\begin{tabular}[c]{@{}l@{}}diabeteshormonecenter.com/\end{tabular} & \begin{tabular}[c]{@{}l@{}}DHC\end{tabular} & Website & Generic  & Hospital \\ \hline
\begin{tabular}[c]{@{}l@{}}ceylincocancercentre.lk/\end{tabular} & \begin{tabular}[c]{@{}l@{}}Ceylinco\end{tabular} & Website & Generic  & Medical Services \\ \hline
\begin{tabular}[c]{@{}l@{}}cdem.lk/\end{tabular} & \begin{tabular}[c]{@{}l@{}}CDEM\end{tabular} & Website & Generic  & Medical Services \\ \hline
\begin{tabular}[c]{@{}l@{}}com.health\_assist\end{tabular} & \begin{tabular}[c]{@{}l@{}}Nawaloka\end{tabular} & Android App & Generic  & Hospital \\ \hline
\begin{tabular}[c]{@{}l@{}}com.durdans.patientcare\end{tabular} & \begin{tabular}[c]{@{}l@{}}Durdans\end{tabular} & Android App & Generic  & Hospital \\ \hline
\begin{tabular}[c]{@{}l@{}}www.healthguard.lk/\end{tabular} & \begin{tabular}[c]{@{}l@{}}Health Guard\end{tabular} & Website & Generic  & Pharmacy \\ \hline
\begin{tabular}[c]{@{}l@{}}unionchemistspharmacy.lk/\end{tabular} & \begin{tabular}[c]{@{}l@{}}Union Chemist\end{tabular} & Website & Generic  & Pharmacy \\ \hline
\begin{tabular}[c]{@{}l@{}}www.mycare.lk/\end{tabular} & \begin{tabular}[c]{@{}l@{}}mycare\end{tabular} & Website & Generic  & Pharmacy \\ \hline
\begin{tabular}[c]{@{}l@{}}www.uniquepharmacy.lk/\end{tabular} & \begin{tabular}[c]{@{}l@{}}Unique Pharmacy\end{tabular} & Website & Generic  & Pharmacy \\ \hline
\begin{tabular}[c]{@{}l@{}}www.carelink.lk/\end{tabular} & \begin{tabular}[c]{@{}l@{}}Carelink\end{tabular} & Website & Generic  & Pharmacy \\ \hline

\begin{tabular}[c]{@{}l@{}}www.ceymed.lk/\end{tabular} & \begin{tabular}[c]{@{}l@{}}Ceymed\end{tabular} & Website & Generic  & Medical Service \\ \hline
\begin{tabular}[c]{@{}l@{}}buymedicine.lk/\end{tabular} & \begin{tabular}[c]{@{}l@{}} Buy Medicine \end{tabular} & Website & Generic  & Pharmacy \\ \hline
\begin{tabular}[c]{@{}l@{}}onlinepharmacy.lk/\end{tabular} & \begin{tabular}[c]{@{}l@{}}Online Pharmacy\end{tabular} & Website & Generic  & Pharmacy \\ \hline
\begin{tabular}[c]{@{}l@{}}echannelling.com/\end{tabular} & \begin{tabular}[c]{@{}l@{}}eChanneling \end{tabular} & Website & Physician Appointment  & Third-Party \\ \hline
\begin{tabular}[c]{@{}l@{}}vida.lk\end{tabular} & \begin{tabular}[c]{@{}l@{}}Vida\end{tabular} & Website & Generic  & Medical Services \\ \hline
\begin{tabular}[c]{@{}l@{}}www.odoc.life/\end{tabular} & \begin{tabular}[c]{@{}l@{}}oDoc - \\Video Consultations\end{tabular} & Website & Generic  & Medical Services \\ \hline

\end{tabular}%
\caption{List of applications tested}
\label{attribution}
\end{table*}

\end{document}